\documentclass[%
 preprint,
 amsmath,amssymb,
 aps,
]{revtex4-2}

\usepackage{float}
\usepackage{graphicx}
\usepackage{dcolumn}
\usepackage{bm}
\usepackage{balance}

\begin{document}

\title{Template-Free Fabrication
of Reconfigurable Magnetic Micropillars and Filaments
through Controlled Nanoflower Assembly and Actuation}

\author{Caterina Landi}%
\altaffiliation{These authors contributed equally.}%
\affiliation{%
  Departamento de Estructura de la Materia, F\'isica T\'ermica y Electr\'onica, 
  Universidad Complutense de Madrid, 28040 Madrid, Spain.
}%

\author{Rosa P\'erez-Garrido}%
\altaffiliation{These authors contributed equally.}%
\affiliation{%
  Departamento de Qu\'imica F\'isica, Facultad de Ciencias Qu\'imicas, 
  Universidad Complutense de Madrid, Ciudad Universitaria s/n, 28040 Madrid, Spain.
}%

\author{Julio Marco Cuenca}%
\affiliation{%
  Departamento de Estructura de la Materia, F\'isica T\'ermica y Electr\'onica, 
  Universidad Complutense de Madrid, 28040 Madrid, Spain.
}%

\author{Javier Tajuelo}%
\affiliation{%
  Department of Interdisciplinary Physics, Universidad Nacional de Educacion 
  a Distancia UNED, Avda. Esparta s/n, 28232 Las Rozas, Spain.
}%

\author{Chantal Valeriani}%
\affiliation{%
  Departamento de Estructura de la Materia, F\'isica T\'ermica y Electr\'onica, 
  Universidad Complutense de Madrid, 28040 Madrid, Spain.
}%

\author{Helena Gavil\'an}%
\email{hgavilan@ucm.es}%
\affiliation{%
  Departamento de Qu\'imica F\'isica, Facultad de Ciencias Qu\'imicas, 
  Universidad Complutense de Madrid, Ciudad Universitaria s/n, 28040 Madrid, Spain.
}%

\author{Fernando Mart\'inez-Pedrero}%
\email{fmpedrero@ucm.es}%
\affiliation{%
  Departamento de Qu\'imica F\'isica, Facultad de Ciencias Qu\'imicas, 
  Universidad Complutense de Madrid, Ciudad Universitaria s/n, 28040 Madrid, Spain.
}%

\date{\today}
\maketitle
             
\section{Introduction}

Often inspired by biological systems, researchers have fabricated a wide variety of advanced materials with complex nano-structures and functional sophistication. In such materials, key properties —including electrical conductivity, permeability, and magnetic responsiveness— can be finely tuned by manipulating parameters such as nanoparticle size, surface functionalization, and inter-particle spacing \cite{Ashby2025, Whitesides2002}. In this context, magnetic nanoparticles (MNPs) have attracted significant attention due to their ability to form responsive, field-directed static and dynamic self-assemblies. Among them, magnetic nanoflowers (MNFs) represent a particularly appealing platform, owing to their distinctive magnetic properties. These features have enabled their widespread use in diverse applications, including magnetic hyperthermia, biomedical imaging, targeted drug delivery, and water treatment \cite{Lartigue2012, GarciaSoriano2024, GalloCordova2022, Rosensweig2013}.
In magnetic colloids, controlled application of magnetic fields enables precise tuning of microstructural features, which in turn govern the structural and functional properties of the resulting materials \cite{Rosensweig2013}. The choice of material components—ranging from purely inorganic to polymer-based composites or biohybrids—plays a critical role in determining attributes such as flexibility, biocompatibility, and magnetic responsiveness, and is typically tailored to meet the demands of specific applications \cite{Wei2024}. Across all material platforms, the assembly process must strike a delicate balance: magnetic interactions, which drive particle self-assembling, compete with other colloidal forces and thermal fluctuations, which might undermine structural integrity.
Among various nanocomposite architectures, colloidal magnetic micropillars -elongated structures anchored to a substrate- and microfilaments -free-standing- have attracted significant interest as advanced functional materials, owing to their unique properties arising from their quasi-one-dimensional geometry and intrinsic magnetism \cite{Ni2023, Kwon2024, Bharti2015}. Micropillars are widely employed in surface engineering, sensing, and photonics, whereas detached microfilaments find applications in soft robotics, actuation, and biomedical technologies \cite{Ni2023, Belardi2011, Vilfan2010}.
By adjusting a magnetic field, one can change the orientation of many filaments and thereby change the material's properties, such as optical transparency, wettability or mechanical stiffness, in real time \cite{Doyle2002, Nie2010, Fiser2015, Wang2019, Luo2019}. On the other hand, flexible magnetic filaments can undulate like a tiny whip under an alternating magnetic field. This flexibility not only enables biomimetic movement but also allows magnetic microswimmers to perform oscillatory, non-reciprocal actuations—essential for achieving net motion at low Reynolds numbers \cite{Dreyfus2005}. In soft robotics, larger assemblies of magnetic filaments or bundles can serve as tentacles, grippers, or limbs that bend on demand \cite{Kim2022}. Rigid magnetic filaments have also been employed as rheological probes in interfacial systems. Their quasi-one-dimensional geometry and high aspect ratio provide an extended interfacial contact area, thereby enhancing sensitivity to subtle viscoelastic responses at fluid interfaces \cite{Tajuelo2016}. Magnetic filaments are also gaining attention for applications in targeted drug delivery and minimally invasive medical interventions \cite{Cebers2016, Mateos2019}, and in magnetic hyperthermia, being able to dissipate heat under high frequency alternating magnetic fields \cite{Alphandery2012, Alexandridis2025}.
Magnetic micropillars are typically fabricated through template-assisted synthesis, using porous molds such as anodic aluminum oxide (AAO) membranes, where magnetic materials are deposited or self-assembled within the pores \cite{Fiser2015}. Other microfabrication techniques include soft lithography, photolithographic patterning, and related methods \cite{Belardi2011, Vilfan2010}. In the first case, anchored micropillars can be released and converted into microfilaments once their formation is complete, removing the mold to liberate the structures through the chemical dissolution of the template \cite{Ni2023, Zamani2022, Wang2003}.
On the other hand, magnetic filaments can also be produced via template-free methods. In these approaches, external magnetic fields induce dipole moments in the constituent particles, promoting their alignment into head-to-tail chains which can further assemble into bundled configurations. However, since particles are typically superparamagnetic, the resulting structures are generally unstable and prone to disintegration due to thermal agitation once the external field is removed. To achieve structural permanence, a secondary mechanism, such as polymerization, silica deposition or capillary bridging, is often required to bind the particles together and form stable structures \cite{Golovanov2013, Vereda2007, Kralj2015}. Other related bottom-up methods include electrostatic co-assembly \cite{Golovanov2013}, or field-induced self-association in the presence of block copolymers \cite{FrkaPetesic2011}. In these processes, tuning the electrostatic interactions—primarily through modulation of the dispersion's ionic strength, enables the formation of micrometer-scale spherical or elongated aggregates \cite{Fresnais2008}. The synthesis of magnetoresponsive bonded chains can also be achieved by linking paramagnetic microparticles \cite{Dreyfus2005, Biswal2003}. However, all these methods provide only limited control over the size of the resulting self-assembled structures. In contrast, biotemplating employs biological filaments, such as flagella, microtubules, or plant fibers, as structural templates \cite{Magdanz2020, Bereczk2017}.
Alternatively, electrospinning is a versatile technique for producing magnetic micro/nanofibers. In this approach, a charged polymer solution is ejected into a strong electric field, and MNPs can be incorporated into the spinning solution to produce magnetizable nanofibers in a single step. These electrospun polymer fibers can serve as sacrificial templates for subsequent magnetic material deposition \cite{Yusoff2021}.
However, many of these processes offer limited flexibility in design, typically permitting only minor variations in geometry or material composition. Emerging fabrication technologies are expanding the possibility of designing magnetic structures even further. Using magnetically functional inks, 3D printing techniques, now allow the direct writing of complex architectures layer by layer \cite{Kim2022}. Light-based 3D printing techniques, such as stereolithography, can similarly orient and fix magnetic particles within photocurable resins, achieving fine feature sizes on the order of tens of micrometers \cite{Xu2019}. Finally, it is worth noting that magnetic microwires were also prepared by rapid solidification using the quenching and drawing techniques \cite{Zhukov2004, Chiriac2011}.
Building on our previous discussion, both self-assembly strategies and the use of structured templates or molds have traditionally been employed to guide the formation of elongated nanostructured materials, ranging from ordered arrays of micropillars to free-standing microfilaments. While effective, these approaches suffer from several drawbacks: they can compromise monodispersity and flexibility, introduce contaminants during fabrication, face scalability challenges, or show limited applicability in processes that require dynamic adaptability. Building on approaches in which a glass slide is positioned above the substrate to promote the controlled growth of magnetic micropillars with uniform lengths and ordered arrangements \cite{Wang2019, Luo2019}, we have extended these strategies to enable the fabrication of both anchored micropillars and free-standing magnetic filaments.
In this work, we propose a template-free strategy, for fabricating structures with controlled shapes, sizes, and properties. By modulating ionic strength and confinement, controlling MNFs surface charge, and applying external magnetic fields, we present a novel approach in which MNFs suspended in an aqueous solution spontaneously self-organize into micropillars and microfilaments, without the need for predefined scaffolds. The experiments were performed using uncoated MNFs and MNFs coated with L-dopamine (MNFs@L-DOPA). L-DOPA (L-3,4-dihydroxyphenylalanine) was selected because it is a biocompatible small molecule with an established clinical safety profile, which may facilitate future applications in targeted drug delivery and minimally invasive actuation-based medical interventions. Owing to its catechol, amine, and carboxylate functionalities, L-DOPA provides a net negative surface charge and can form a thin protective layer on the nanoparticle surface while simultaneously offering reactive groups for biomolecular coupling, without the need for polymeric coatings \cite{Youshia2024, Kalcec2022}. The resulting assemblies exhibit either reversible behavior or structural stability under controlled conditions and can be externally actuated to generate dynamic motility.

\section{Materials and Experimental Methods}

\subsection{Materials}

Ferric chloride hexahydrate (FeCl$_3$·6H$_2$O, 98\%, Sigma-Aldrich), ferrous chloride tetrahydrate (FeCl$_2$·4H$_2$O, 98\%, Sigma-Aldrich), Ferric nitrate nonahydrate (Fe(NO$_3$)$_3$·9H$_2$O, 98\%, Thermo-Scientific), sodium hydroxide (NaOH, 98\%, Sigma-Aldrich), sodium chloride (ACS reagent, $\geq$99.0\%, Sigma-Aldrich), hydrochloric acid (HCl, 37\%, Scharlab), (HNO$_3$, 60\% w/w; Scharlab), N-methyldiethanolamine (NMDEA, 99\%, Sigma-Aldrich), diethylene glycol (DEG, 99\%, Sigma-Aldrich), ethyl acetate (99.8\%, Thermo-Scientific), ethanol (96\%; Scharlau), acetone (for analysis, Thermo Scientific Chemicals) 3,4-Dihydroxy-L-phenylalanine (L-DOPA 98\%, Thermo-Scientific), Tris(hydroxymethyl)aminomethane (Tris/THAM, Certified ACS, Fisher-Scientific) and Milli-Q (18.2 M$\Omega$·cm) water were used in this study. Anionic surfactant sodium dodecyl sulfate (SDS, $\geq$99.0\%, SigmaAldrich) was employed for post-field treatment and surface functionalization of assembled structures. Solvents used in this work were used as received, without further purification.

\subsection{Polyol synthesis of magnetic iron oxide nanoflowers}

The synthesis of MNFs was adapted from \cite{GalloCordova2022, Hugounenq2012, Gavilan2017}. Briefly, 0.203 g of iron (II) and 0.800 g of iron (III) salts were dissolved in 32.0 g of DEG and 32.0 g of NMDEA. Separately, 0.512 g of NaOH was dissolved in 16.0 g of DEG and 16.0 g of NMDEA by ultrasonication at 60 °C for 90 min. Both solutions were briefly stirred magnetically ($\sim$5 min), then transferred into a Teflon-lined stainless-steel autoclave, sealed, and heated in a preheated oven at 220 °C for 10 h. The nanoflowers were then subjected to an acid treatment \cite{Costo2012}: first resuspended in 2.0 M HNO$_3$ under vigorous magnetic stirring for 15 min, followed by redispersion in an aqueous solution of iron (III) nitrate (8.2 g in water) and heating at 90 °C for 30 min with stirring. After cooling to room temperature, the MNFs were washed with 2.0 M nitric acid, then washed three times with ethanol and acetone, and finally redispersed in deionized water. Any remaining acetone was removed using a rotary evaporator, and the final MNFs suspension was prepared in 10 mL of deionized water.

\subsection{L-DOPA coating procedure}

The coating procedure was based on a protocol previously reported by some of us \cite{Mandriota2025}, with minor modifications. Briefly, a 1.0 M Tris (THAM) solution was prepared and adjusted to physiological pH by addition of HCl under pH-meter control, then diluted with Milli-Q water to obtain 10 mM Tris buffer (pH 7.4). MNFs stock dispersion (typically ca. 10-20 mg mL$^{-1}$) was diluted with Milli-Q water to a working concentration of 3 mg mL$^{-1}$ prior to coating. L-DOPA was dissolved in 10 mM Tris buffer (pH 7.4) at concentrations of 25--65 mM with brief sonication ($\sim$5 min), and the pH was adjusted to 8.5 using NaOH to initiate oxidative processes of L-DOPA leading to surface coating formation. Aliquots of the MNFs dispersion (200 µL, 3 mg mL$^{-1}$) were mixed with L-DOPA solution (300 µL) in microcentrifuge tubes and incubated on an orbital shaker at room temperature for 3 h, enabling in situ surface modification of the MNFs using L-DOPA. The resulting L-DOPA--modified MNFs were washed three times with Milli-Q water by magnetic separation (typically complete magnetic separation occurs after 1-2 h). The procedure is readily scalable by a factor of 10--25. L-DOPA-stabilized nanoflowers were redispersed in Milli-Q water to the required concentration and stored at 4 °C for further assembly experiments. The measured zeta potential was ($-36 \pm 4$) mV, indicating excellent colloidal stability through electrostatic repulsion arising from deprotonated carboxylate groups of the L-DOPA surface modification.

\subsection{MNFs characterization}

For magnetization measurements using direct current (DC) magnetometry (MPMS-XL magnetometer (Quantum Design) equipped with a SQUID (Superconducting Quantum Interference Device) detector), a defined volume (60 µL) of MNF aqueous solution with an iron concentration of 3 mg/mL, as determined by inductively coupled plasma optical emission spectroscopy (ICP-OES, Spectro Arcos III), was drop-cast onto cotton wool, which served as both sample holder and substrate. After a mild drying process, 60 ºC for 8 h in a furnace, the cotton wool is placed in a polycarbonate capsule, and DC magnetization curves were registered at room temperature ($T = 298$ K).

The magnetic heating performance of the iron-oxide nanoflowers was evaluated by AC magnetic hysteresis loop measurements, which provide a direct and calibration-free determination of the specific absorption rate (SAR). Measurements were performed using an induction magnetometer (AC-Hyster, Nanotech Solutions) under alternating magnetic fields with amplitudes $H =$ 8, 12, 16, 20, and 24 kA·m$^{-1}$ and excitation frequencies $f =$ 100, 200, and 300 kHz, at a fixed iron concentration of [Fe] = ca. 1.5 mg·mL$^{-1}$ (determined through ICP-OES). Magnetic hyperthermal response was investigated by recording hysteresis loops for the same. SAR values were calculated directly from the hysteresis loop area according to SAR = $(f /\rho_\mathrm{Fe}) \oint M(H)\mathrm{d}H$.

MNFs were dispersed in NaCl solutions at the desired ionic strengths and thoroughly sonicated to ensure homogeneous suspensions. The nanoparticle number concentration of the resulting suspensions, [MNFs] $= 1.83 \times 10^{13}$ nanoparticles/mL, 30.4 nM, was calculated based on the density of Fe$_3$O$_4$, assuming spherical particle geometry and an average diameter of 28 nm. Sodium chloride (NaCl) solutions were prepared at concentrations ranging from 0.1 mM to 100 mM to modulate the ionic strength of the medium. Increasing the electrolyte concentration lowers the measured zeta potential and screens the electrostatic repulsion between particles. As will be discussed in the results and is well documented in the literature \cite{Bacri1994, Bacri1989, Dubois2000, Cousin2001, Raboisson2020, Erdmanis2017}, the reduction in electrostatic repulsion, combined with magnetic interactions arising from the weakly ferromagnetic character of the MNFs, promotes phase separation of the suspension into liquid-like and gas-like domains at moderate salt concentrations, while inducing the formation of irreversible, solid-like amorphous aggregates at higher salt concentrations. Both processes can be accelerated under the influence of an external magnetic field \cite{Bacri1994, Bacri1989, Dubois2000, Cousin2001, Raboisson2020, Erdmanis2017}.

The suspension is loaded by capillary action into a rectangular chamber formed between a standard microscope glass slide (RS) and a coverslip (Menzel-Gläser). The chamber height is defined by two parallel strips of rectangular adhesive tape, which act as spacers and create an internal volume measuring approximately 1.0 cm wide, 2.0 cm long, and $\sim$50 µm high. Magnetic fields were applied in static, oscillatory, or rotating modes, in both vertical and horizontal configurations, to induce assembly. The open-source image processing software ImageJ (National Institutes of Health) has been essential for determining structural characteristics from optical and electronic microscopy images. These tools enabled quantification of inter-columnar distances, filament lengths and widths, and motile behavior under varying field conditions. Anionic surfactant sodium dodecyl sulfate (SDS) was employed for post-field treatment and surface functionalization of assembled structures.

Real-time observations of MNFs behavior under applied fields were conducted using optical microscopy (Olympus BH2 with a 20$\times$0.25 NA objective). In these experiments, magnetic fields were generated using a custom-built Helmholtz coil system comprising two amplifiers (KEPCO, bipolar power supply/operating amplifier, USA) and five coils that generate controlled magnetic fields along the three spatial directions (four coils control the magnetic field within the horizontal plane and one coil along the vertical axis). A custom-made LabVIEW software was developed to, first, control the amplifiers and, second, monitor the actual current in the coils through an NI-9269 and an NI-9215 DAQ boards, respectively. Given the size and impedance of the coils, the amplitude and frequency of the generated magnetic fields can reach up to 5.0 mT and 100.0 Hz, respectively. A proper synchronization of the three currents enables the generation of static, oscillatory, and rotating magnetic fields in any direction and, in general, any Lissajous figure in 3D enabling precise control over the assembly conditions. Structural characterization of the assembled formations was performed using scanning electron microscopy (SEM, JEOL 6400 JSM), to resolve morphology at micro- and nanoscales. Dynamic light scattering (DLS) using a Zetasizer and ALV system (Malvern) was also employed to measure particle size distribution and zeta potential.

\section{Results and Discussions}

Uncoated magnetic nanoflowers (MNFs) were used as building blocks for field-driven
self-assembly. TEM images exhibited a broad size distribution with a mean Feret
diameter of 30.5 $\pm$ 0.9 nm (Figure S1, A and B). TEM also confirmed their characteristic
multicore, flower-like morphology composed of closely packed nanocrystals, leading to
strong dipolar and exchange interactions and collective superferromagnetic behavior at
room temperature \cite{Hugounenq2012, Kimmel2019}. DC magnetization measurements revealed ferromagnetic-like
hysteresis with $M_s$ (saturation magnetization) = 10$^3$ A·m$^2$·kg$^{-1}$, $M_r$ (remanent
magnetization) = 12 A·m$^2$·kg$^{-1}$, and $H_c$ (coercive field) = 19 kA·m$^{-1}$, which is essential for
dense phase formation and field-induced structures stable even after field removal
(Figure S1, C). Uncoated MNFs tend to interact during TEM sample preparation, likely
due to drying effects, which makes the visualization of individual MNFs difficult. In
contrast, TEM analysis after adsorption of TMAOH molecules on the MNF surface
reveals that the magnetic nanoflowers are composed of a small number of fused
nanocrystallites (``petals'') forming a single flower-like entity with a monodisperse size
distribution (see Figure S1, D--E).

Loop area and SAR increasing nonlinearly with field amplitude and frequency, reaching
up to 575 W g$^{-1}$ at 300 kHz and 24 kA m$^{-1}$, confirming that heating efficiency arises from
intrinsic hysteretic losses associated with the strongly coupled multicore architecture
rather than relaxation-dominated single-core behavior (Figure S2). On the other hand,
DLS measurements at pH 2 showed a hydrodynamic diameter of 70 $\pm$ 11 nm (PdI = 0.3),
with no significant aggregation above 100 nm, and a positive zeta potential of 30 $\pm$ 1 mV,
indicating good colloidal stability despite the absence of surface coating.

\subsection{Behavior of the suspension of uncoated particles at different electrolyte concentrations, in the absence and presence of a magnetic field}

Since the glass walls of the measurement chamber are negatively charged and the
uncoated nanoparticles carry a net positive charge at pH 2.0, particles tend to physisorb onto the surfaces, forming a thin, firmly anchored layer, visible in the microscopy images.
The remaining nanoparticles that do not adhere to the walls stay suspended in the medium, generating a visible ``visual noise'' under optical microscopy, arising from light scattered
by the Brownian nanoparticles (Movie 1). As previously mentioned, increasing the NaCl concentration promotes the screening of the MNFs surface charge, reducing electrostatic repulsion between particles. This, combined with the magnetic interactions exhibited by the ferromagnetic particles---even in the absence of an external magnetic field--- facilitates phase separation within the initially homogeneous suspension. When the ionic strength is increased, MNFs not adsorbed onto the glass surface redistribute into two distinct phases: a dense, liquid-like phase that settles at the bottom of the chamber, and a
dilute, gas-like phase that occupies the remaining volume \cite{Bacri1994, Bacri1989, Dubois2000, Cousin2001, Raboisson2020}. At the interface
between the two phases, a significant interfacial tension, on the order of 10$^{-3}$ mN/m \cite{Bacri1989},
drives the system to minimize the interface area. This results in the formation of droplet-
and groove-like structures of the dense phase, which can respond and deform under the
influence of an external magnetic field \cite{Erdmanis2017, Rowghanian2016}.

Figure 1 (top row) illustrates in detail the evolution of this phase separation across a
range of NaCl concentrations between 0.0 and 100.0 mM. At low salt concentrations (0.1--
50.0 mM), the suspended nanoparticles form small, dense domains that rest on the
substrate. Within these domains, particle diffusion is still observable. Between (50.0-60.0
mM), a more extensive dense phase appears at the bottom of the chamber, adopting
ramified or cracked geometries, whereas at higher salt concentrations, 60.0-70.0 mM, the
system transitions once more into isolated, droplet-like aggregates (results not shown).
The reduction in the extent of the dense phase may result from an increased number of
nanoparticles aggregating and adsorbing onto the bottom surface as the ionic strength
rises. This adsorption decreases the population of freely suspended particles, thereby
limiting the formation of larger, interconnected dense domains. At higher NaCl
concentrations ($>$70.0 mM), the increased ionic strength results in near-complete
screening of surface charges, allowing van der Waals attractions to dominate. Therefore,
the MNFs undergo strong and irreversible aggregation, adhering both to each other and
to the glass walls. This immobilization suppresses Brownian motion, rendering it
unobservable under the microscope.

\begin{figure}[h]
    \centering
    \includegraphics[width=\textwidth]{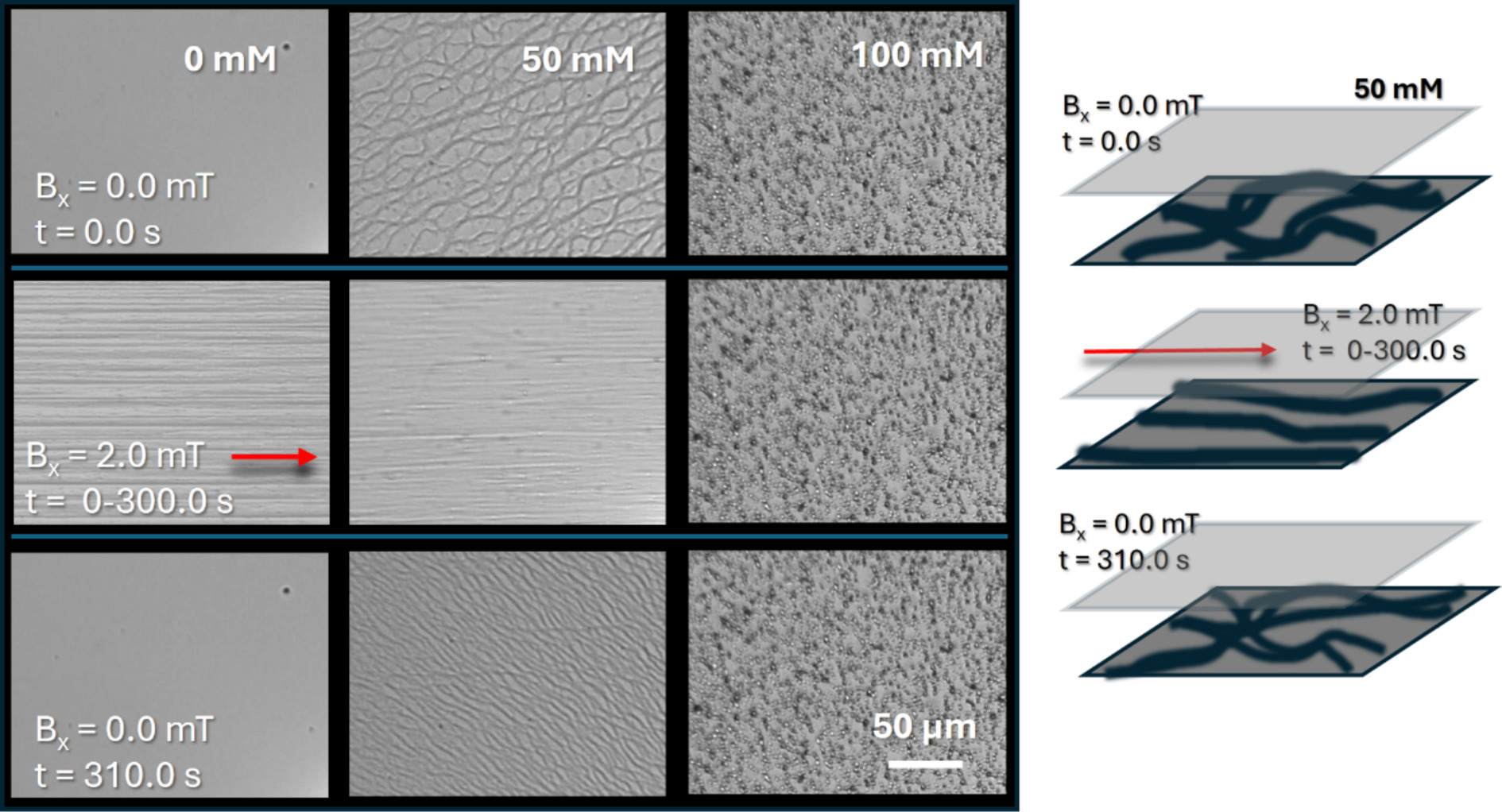}
    \caption{Aqueous suspension of uncoated MNFs at pH = 2.0 with varying NaCl concentrations under a
    constant external magnetic field, applied horizontally, parallel to the glass substrate and perpendicular to
    the microscope's optical axis. Top row: Colloidal suspension prior to magnetic field application. Middle
    row: Suspension under a magnetic field ($B_x = 2.0$ mT) for 300 s. Bottom row: Suspension following field
    removal (Movie 1). The scheme on the right represents the field-induced process in 3D at [NaCl] = 50 mM.}
    \label{fig:Fig1}
\end{figure}

Next, the response of these systems to the application of an external magnetic field was
explored, examining their behavior under different field configurations. Since colloidal
agglomeration arising from partial screening of electrostatic interactions can interfere
with field-induced aggregation, the sample history is a critical factor \cite{Raboisson2020}. Therefore, all
experiments were conducted within minutes of thoroughly mixing the particle suspension
with the electrolyte solution. When a magnetic field is applied parallel to the substrate,
i.e., perpendicular to the optical axis of the microscope, the suspended MNFs display a
strong tendency to form linear aggregates and columnar structures, provided they remain
mobile, that is, at NaCl concentrations below 80.0 mM (Figure 1, middle row). This
column formation arises from the alignment of magnetic dipoles induced by the applied
field, as well as from attractive magnetic lateral interactions between emerging chains of
particles under the influence of the external field \cite{MartinezPedrero2020}. The column formation process
occurs rapidly upon field application, typically stabilizing within a few seconds. This
swift response indicates that, at the field strengths used, field-induced dipole--dipole
interactions dominate over thermal motion. Once the magnetic field is turned off, the
columns disassemble and the nanoparticles redisperse into the dense phase, independently
of the salt concentration, as the field-induced magnetic interactions driving the columnar
assembly are transient and remain active only in the presence of the external field (Movie
1, Figure 1, bottom row).

\begin{figure}[h!]
    \centering
    \includegraphics[width=\textwidth]{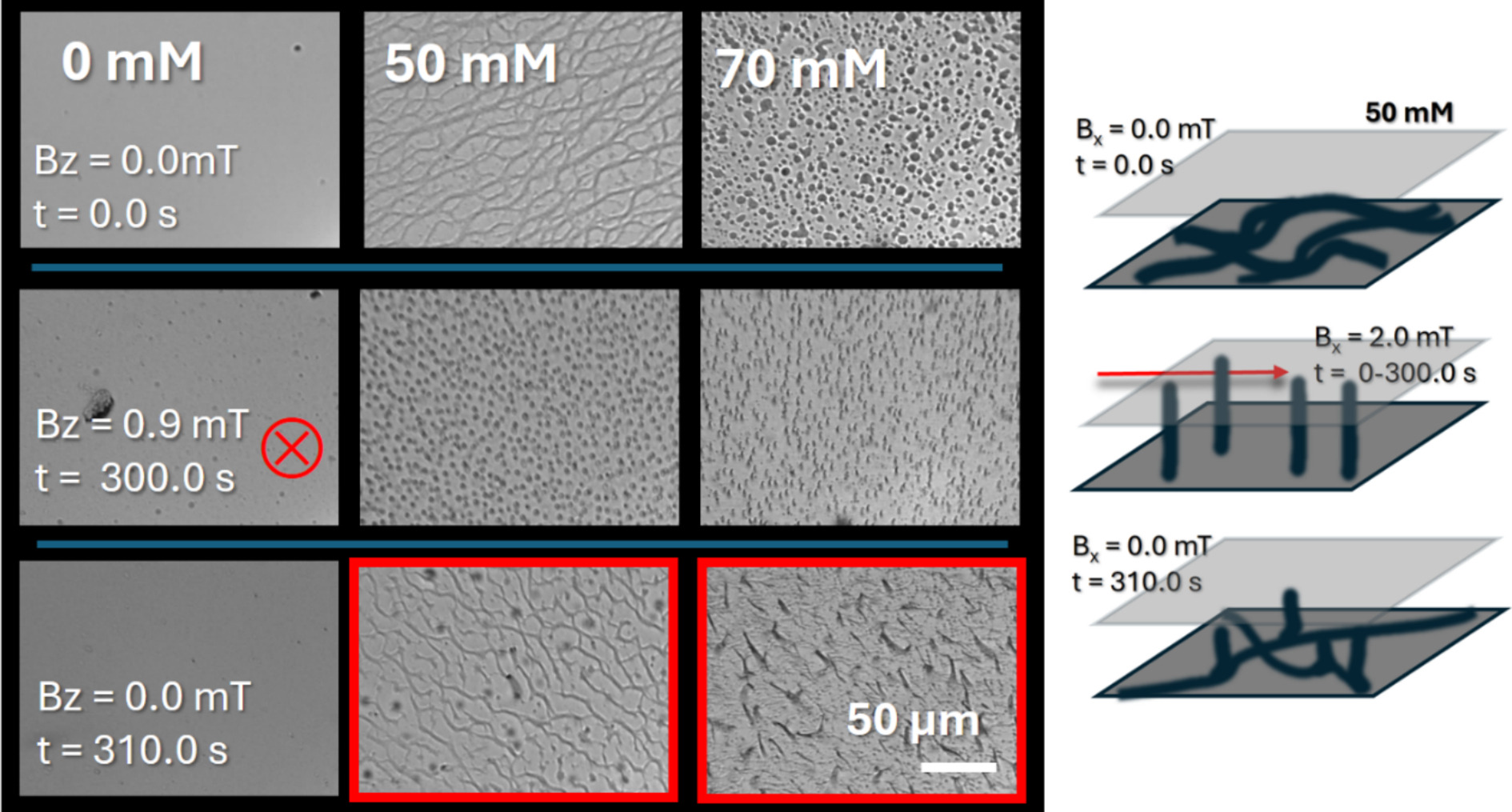}
    \caption{Suspension of uncoated MNFs at pH = 2.0 with different NaCl concentrations under a constant
    external magnetic field oriented vertically, perpendicular to the substrate plane. Top row: Colloidal
    suspension prior to magnetic field application. Middle row: Suspension under a magnetic field ($B_z = 0.9$
    mT) for 300 s. Bottom row: Suspension following field removal. Red-framed images highlight conditions
    in which the filaments formed under the action of the field remain visible for seconds after the field is
    turned off (Movie 2). The scheme on the right represents the field-induced process in 3D at [NaCl] = 50
    mM.}
    \label{fig:Fig2}
\end{figure}

When the magnetic field is applied perpendicular to the interface, aligned with the vertical
axis, the behavior of the suspension changes markedly. The field induces the formation
of columnar structures that grow vertically from the dense phase, resting at the bottom of
the chamber (Figure 2, middle row). At the particle concentrations studied, these
columns remain spatially separated and are distributed across the field of view without
exhibiting long-range positional or orientational order. The average distance between
columns is on the order of several micrometers, suggesting a spacing mechanism
governed by repulsive interactions, likely dominated by field-induced dipole--dipole
forces. As will be discussed later, the columns present a uniform width of several
micrometers, and approximately one-third of them span the entire height of the container,
as confirmed by adjusting the microscope's focal plane.

At NaCl concentrations below 50.0 mM, the columnar structures are fully reversible and
reintegrate into the dense phase upon removal of the magnetic field. At intermediate salt
concentrations (50.0--65.0 mM), the process remains reversible, with the structures
disassembling once the field is switched off. However, the reintegration of particles into
the dense phase becomes noticeably slower (Figure 3, Movie 2). The characteristic decay
time of reintegration, $t_c$, increases exponentially with electrolyte concentration, indicating
that higher ionic strengths slow down the relaxation dynamics of the system.

\begin{figure}[h!]
    \centering
    \includegraphics[width=\textwidth]{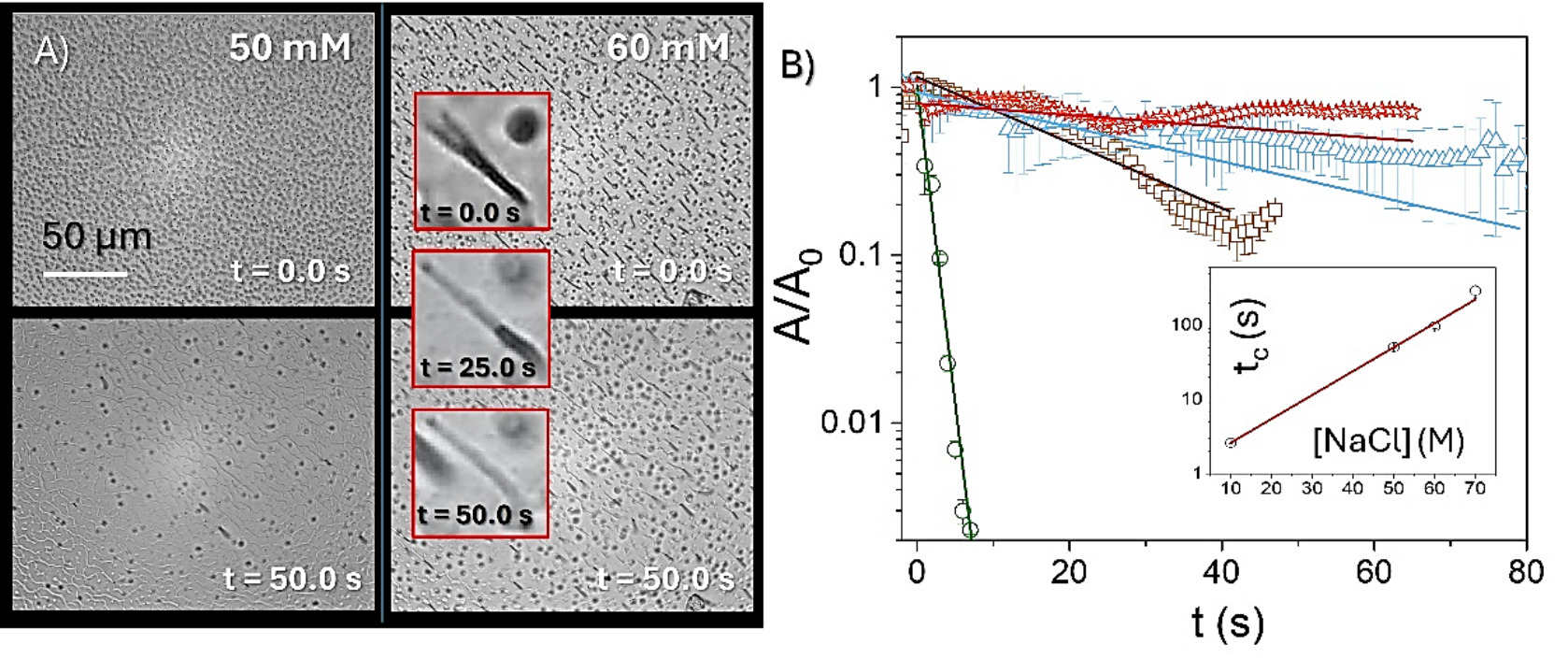}
    \caption{A) Columnar formations of MNFs under the application of a constant vertical magnetic field ($B_z
    = 0.9$ mT) at pH = 2.0 with two different NaCl concentrations. Top row: Structures observed under the
    applied field for 300 s. Bottom row: Same regions observed after field removal (Movie 2). Framed insets
    show the time evolution of a representative filament formed at [NaCl] = 60 mM. B) Logarithm of the
    projected area by the filaments, normalized to its value just after the field along the z-axis was switched
    off, $A/A_0$, plotted as a function of time. The averaged values and associated errors result from the analysis
    of three different micropillars monitored under the same conditions. The inset shows the linear relationship
    between the logarithm of the characteristic absorption time, $t_c$, and [NaCl].}
    \label{fig:Fig3}
\end{figure}

For salt concentrations in the range of 70.0--80.0 mM, the system transitions from
reversible to irreversible behavior, and the filaments no longer reintegrate into the dense
phase---at least not within timescales of several minutes. This points to the predominant
role of van der Waals, which become increasingly significant as electrostatic repulsion is
further suppressed, thereby stabilizing the formed filaments \cite{Bharti2015, MartinezPedrero2008}. It is also noteworthy
that, after the application of the vertical field, a portion of these irreversible filaments
become anchored to the upper surface of the container, as can be confirmed by adjusting
the microscope's focal plane. Above 90.0 mM, the system ceases to respond to the vertical
magnetic field. As commented before, the near-complete suppression of electrostatic
repulsion leads to strong, irreversible aggregation of the nanoparticles, which remain
immobilized on the glass substrate and exhibit no response to the external field. As shown in Figure S3, the size and spatial distribution of the columns formed at low and
intermediate electrolyte concentrations are strongly influenced not only by particle and
salt concentrations, or the intensity of the applied magnetic field, but also by the height
of the sample chamber. The chamber height plays a critical role in constraining the
vertical extent of column development \cite{Wang2019, Luo2019}.

\subsection{Response of magnetic micropillars to different magnetic field configurations}

The micropillars formed from uncoated particles at salt concentrations in the range of
50.0--90.0 mM, under a vertical external magnetic field tend to anchor to one of the glass
surfaces of the chamber, while still retaining the ability to respond to external magnetic
fields. In the following paragraphs, we will examine their behavior under different
magnetic field configurations.

When a magnetic field is applied parallel to the horizontal plane, the irreversible
micropillars, previously formed under a vertical field and moderate salt concentrations,
reorient along the direction of the new field and gradually aggregate laterally, forming
columns or bundles (Figure 4A, Movie 3). These horizontally aligned structures could
potentially serve as rails for transporting larger particles, such as nano/microparticles or
cells, through magnetically induced directed motion or by the generated hydrodynamic
flow, respectively \cite{MartinRoca2023, MartinezPedrero2021}. When a constant vertical magnetic field is combined with an
oscillating field oriented along the x-axis (Figure 4B, Movie 4), or with a circular rotating
field around the vertical axis (Figure 4C, Movie 5), the irreversible structures previously
formed under the vertical field exhibit dynamic, coordinated motion reminiscent of cilia
\cite{Hamilton2021}. Video microscopy images reveal that at low frequencies, most micropillars oscillate
or precess in phase, moving cohesively in a common direction. However, a smaller
fraction of structures seems to exhibit a 180-degree phase lag in their oscillation or
rotation, respectively. This behavior is attributed to a subset of micropillars being
anchored to the upper surface of the chamber, as previously discussed. Such anchoring
introduces an asymmetry in their response to the external field. This difference is
illustrated in Figures 5B and 5C, where the contrasting dynamics are highlighted with red
and blue lines.

When a rotating magnetic field is applied parallel to the horizontal plane, the magnetic
micropillars formed by the nanoparticles rotate synchronously with the field at low
frequencies (Movie 6). At low Reynolds number, the balance between the magnetic
torque and the viscous friction with the medium and the surface governs the motion of
these rotating micropillars, which remain anchored to the layer of adsorbed nanoparticles
\cite{Tierno2008}. When the magnetic torque, and consequently the angular velocity of the induced
rotation, is moderate, the particles composing the filaments exhibit a higher tendency to
adsorb onto the nanoparticle bed anchored to the bottom glass. This initially promotes the
anchoring of discrete positions along the filament, leading to bending during rotation, and
ultimately to the adhesion of the entire filament to the nanoparticle bed, which
progressively reduces the number of actively rotating structures over time (Figure 4D).
Micropillars formed at low ionic strengths, below 65 mM, gradually bend, stop rotating,
and slowly reintegrate into the dense lower phase. As previously discussed, this behavior
can be attributed to the dominant role of electrostatic interactions, which still stabilize the
colloidal particles and allow them to be reabsorbed into the dense phase.

When the rotating magnetic field is oriented perpendicular to the substrate, $\mathbf{B}(f) =
B_x \sin ft + B_z \cos ft$, with ellipticity $\beta = \frac{B_x^2 - B_z^2}{B_x^2 + B_z^2} \in [-1,1]$, the system's behavior is strongly
influenced by the value of the vertical component of the magnetic field $B_z$. At low field
strengths of the vertical component ($B_z = 0.0$-$0.1$ mT, $B_x = 0.2$ mT, $f = 1.0$ Hz, $\beta \in
[0.6, 1.0]$), the micropillars remain partially attached to the dense lower phase, with
limited ability to move across the surface. At intermediate strengths ($B_z = 0.1$-$0.6$ mT, $B_x
= 0.2$ mT, $f = 1.0$ Hz, $\beta \in [-0.8, 0.6]$), the magnetic torque induced by the rotating field
can occasionally exceed the adhesive force between the micropillar root and the dense
nanoparticle layer on the substrate. As a result, they may detach and begin to move across
the surface, exhibiting direct motion in the plane of field rotation, though asynchronously
with respect to the field. During this locomotion, mobile filaments often encounter
micropillars still anchored to the substrate, with which they tend to aggregate, forming
more complex, crawling-like structures. In some instances, these assemblies extend
protrusions or appendages, adopting shapes and dynamics reminiscent of bacteria or
protozoa (see Figure S4, Movie 7).

\begin{figure}[h!]
    \centering
    \includegraphics[width=\textwidth]{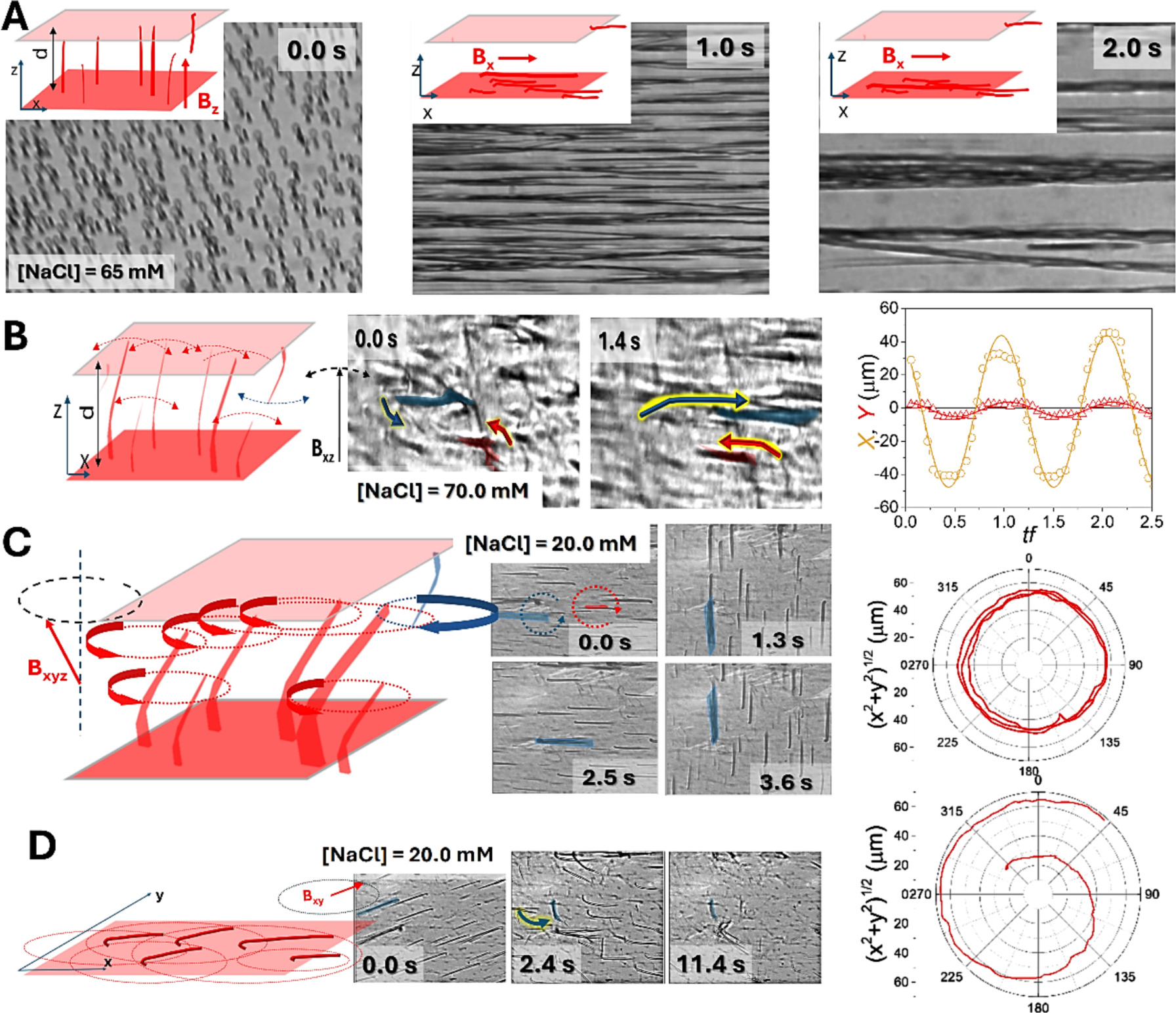}
    \label{fig:Fig4}
\end{figure}

\newpage

\begin{quote}
\textbf{FIG. 4.} 
\textbf{A)} Schematic representation of the strategy for forming irreversible 
parallel tracks adsorbed onto the substrate. Upon applying a constant vertical 
magnetic field, $B_z$, at specific salt concentrations, within the 
65.0--100.0~mM range, irreversible micropillars are formed (left). The resulting 
structures respond to an additional horizontal field, $B_x$ (middle, right). The 
images show micropillar formation under a vertical magnetic field $B_z = 2.0$~mT 
($t = 0.0$~s), alignment of the structures ($t = 1.0$~s), and subsequent lateral 
aggregation of the preformed filaments ($t = 2.0$~s), under a horizontal magnetic 
field $B_x = 1.0$~mT (Movie~3).
\textbf{B)} Schematic representation of the motion followed by the micropillars 
under the oscillating field (left). Their oscillatory motion is indicated by blue 
and red arrows. (middle) Experimental observation of the sample under this field 
configuration ($B_z = 0.9$~mT, $B_x = 2.5$~mT, $f_x = 0.1$~Hz, 
[NaCl] = 70.0~mM, Movie~4). (right) Time evolution of the $x$ and $y$ coordinates 
of the tip position of one of these micropillars over more than two magnetic field 
cycles. The filament marked in red oscillates in antiphase, as it is anchored to 
the upper coverslip.
\textbf{C)} Left: Schematic representation of the motion followed by the 
micropillars under a magnetic field precessing around the vertical axis. Middle: 
Experimental observation of the sample under such a field. The micropillar marked 
in red precesses in antiphase, as it is anchored to the upper glass surface. 
Right: Polar-coordinate evolution of the tip position of one of these micropillars 
over more than two magnetic field cycles (Movie~5. $B_z = 0.9$~mT, 
$B_x = B_y = 2.2$~mT, $f_x = f_y = 0.1$~Hz, [NaCl] = 70.0~mM. The phase shift 
between $B_x$ and $B_y$ is 90$^\circ$).
\textbf{D)} Left: Schematic illustration of filament motion under a rotating 
magnetic field applied parallel to the substrate plane. The rotational motion of 
the micropillars is indicated by red dashed lines. Middle: Time-lapse image 
sequence showing micropillars that initially rotate synchronously with the field 
but gradually bend, cease rotation, and are reabsorbed into the dense phase when 
the process occurs at low salt concentrations ($B_x = B_y = 1.1$~mT, 
$f_x = f_y = 0.2$~Hz, [NaCl] = 60~mM, Movie~6). Right: Polar-coordinate evolution 
of the tip position of one of these micropillars over more than one magnetic field 
cycle.
\end{quote}

Above a certain threshold field intensity ($B_z \geq 0.6$ mT, $B_x = 0.2$ mT, $f = 1.0$ Hz, $\beta \in
[-1.0, -0.8]$), all formed micropillars are peeled from the substrate simultaneously, and
the resulting filaments rotate coherently with the field while ``walking'' along the substrate
in a well-defined direction (Figure 5A and B, Movie 8). The swarm exhibits
omnidirectional steering and high maneuverability. Beyond demonstrating complex
collective behavior, it could enable investigations of fundamental processes in living
systems and serve as a functional bio-microrobotics platform with biomedical potential
\cite{Xu2019swarm}. The synchronous, directional motion arises from the coupling between rotation and
translation, driven by differences in viscous drag at the filament ends \cite{Tierno2008}. Filaments
occasionally and spontaneously switch their direction of motion spontaneously,
introducing additional dynamic complexity into their trajectories. This reversal is
attributed to the inversion of frictional asymmetry at the filament tips when they transition
between the lower and upper surfaces of the chamber. When walking occurs along the
upper wall of the chamber, the direction of motion is reversed (Figure 5C, Movie 9).
Figure 5D shows the dependence of the dimensionless velocity, $\tilde{v} = \frac{v}{2fL}$, on the filament
length, $L$, under a circularly polarized rotating magnetic field ($B_x = B_z = 3.4$ mT, $\beta = 0$),
in semidilute conditions, where the filaments can move freely on the substrate without
interacting or coalescing with other filaments for tens of cycles. At low $L$ and frequency
values, the filaments operate in a synchronous regime, rotating in phase with the field and
walking at maximum efficiency, with $\tilde{v}$ values close to 1. Measured values slightly
greater than 1.0 may result from hydrodynamic effects due to the coordinated motion of
nearby filaments in the sample, or from magnetic interactions with particles adsorbed on
the substrate. As the frequency increases, larger filaments are more likely to transition to
a regime in which they still rotate synchronously with the magnetic field but begin to slip
on the substrate, increasingly alternating between walking along the bottom surface and
the upper wall of the chamber, as previously described. Under the explored experimental
conditions, the observed velocities range from 1 to 350 µm/s, depending on the filament
length and, primarily, the field frequency (Figure S5). At frequencies in the range
between 10.0-100.0 Hz, the particles do not rotate any more synchronously with the
applied field but are still transported on the substrate containing the layer of adsorbed
particles. Here, the more complex mechanism can combine both processes described in
the literature: the hydrodynamic rotation--translation coupling due to the presence of the
substrate \cite{Hamilton2021, Steimel2014}, and the generation of a magnetic traveling potential by the adsorbed
particles, which transports the non-adsorbed particles \cite{MartinRoca2023}.

\begin{figure}[H]
    \centering
    \includegraphics[width=\textwidth]{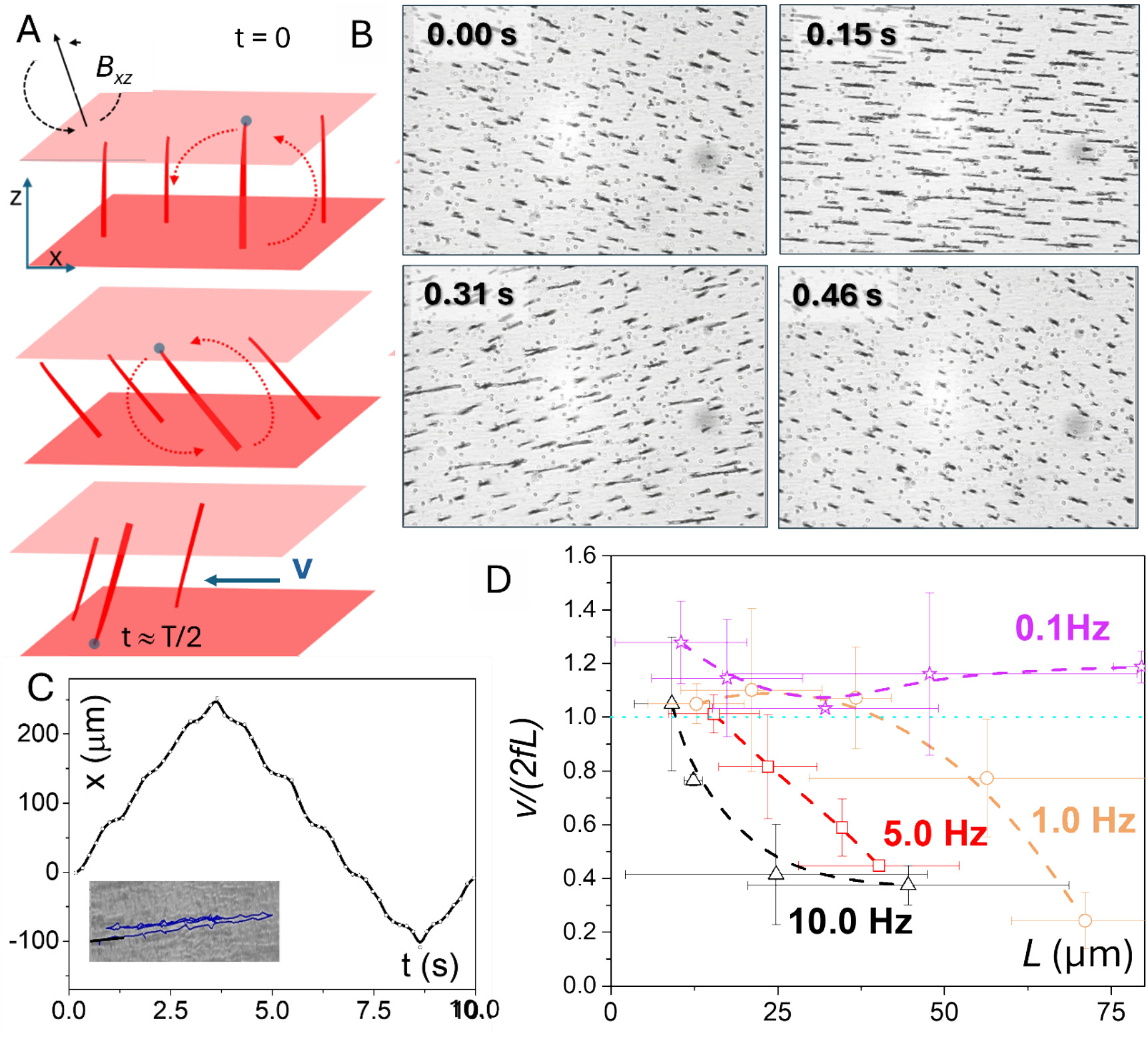}
    \caption{A) Schematic of the transport of magnetic filaments formed under a rotating magnetic field
    applied perpendicular to the substrate plane. The gray spot marks one end of the filament for easy
    identification. B) Sequence of images showing the motion of magnetic filaments under a rotating field
    oriented perpendicular to the substrate ($B_z = 13.5$ mT, $B_x = 3.3$ mT, $f = 1.0$ Hz, $\beta = -0.9$) and at a salt
    concentration of [NaCl] = 70.0 mM. The time of each frame is indicated. C) The image below shows how
    some filaments change their transport direction under the action of the rotating field ($B_x = B_z = 3.3$ mT, $f =
    1.0$ Hz, $\beta = 0$, Movie 9). The graph aside shows the trajectory of a filament along x. D) Dimensionless
    velocity as a function of filament length for different field frequencies, under a circular rotating field applied
    perpendicular to the horizontal substrate ($B_z = B_x = 3.3$ mT, $\beta = 0$). The averaged values and associated
    errors result from the analysis of at least three different filaments of similar length monitored under the
    same conditions.}
    \label{fig:Fig5}
\end{figure}

If the transport strategy is implemented using a rotating field with elliptical polarization,
oriented with its major axis parallel to the solid substrate, the mechanism changes, resembling the motion of a rocking chair that gradually advances through repeated tilting.
At low frequencies, the filament does not complete a full rotation; instead, during the first
quarter of the cycle, it tilts away from the vertical and drags its rear-end forward. During
the second quarter, it reorients back toward the horizontal (Figure S6, Movie 10). When
the filaments are short, they can travel a distance approximately equal to their own length
$L$ during each half-cycle, resulting in a dimensionless velocity $\tilde{v} \approx 1$. In this regime, we
also observe a fraction of filaments moving in the opposite direction. Because the
filaments remain aligned in the same orientation within the substrate plane for longer
periods, interactions and aggregation between filaments become more likely. When the
rotating field is elliptically polarized, with its major axis perpendicular to the solid
substrate, it enhances the desorption of particles, allowing more particles to participate in
the formation of new structures (Figure S7, Movie 11). As previously shown, increasing
the z-component of the field promotes the separation of the field-induced structures.
When the filament density is high, interactions between filaments become significant, and
the velocity appears to decrease. However, it becomes increasingly difficult to define the
filaments as independent entities, since they tend to aggregate with one another during
the application of the rotating field.

\subsection{Peeling of anchored micropillars to harvest size-controlled magnetic filaments}

To extract the filaments from the custom-designed chamber, a low-frequency rotating
magnetic field is applied in a plane perpendicular to the substrate. This couples the
filaments' rotation with translational motion, following the strategy described in the
previous section. At the open ends of the chamber, two droplets of an aqueous solution,
having the same salt concentration as the suspension inside, are placed. Under the action
of the rotating field, the filaments migrate toward one of these open ends and into one of
the droplets. Figure 6A shows the filaments crossing the edge of the coverslip to enter
one of the droplets placed at the side of the chamber (blue shaded area in the scheme).

Once the irreversible filaments---formed at moderate salt concentrations and under
rotating magnetic fields---were extracted from the formation chamber, they were laid
down and aligned under the action of $B_x$. Subsequently, scanning electron microscopy
(SEM) was employed to obtain detailed morphological and structural information on the
resulting structures (Figure 6B). It is important to note that SEM imaging requires prior
drying of the sample, and the formed filaments were able to withstand this process.

\begin{figure}[h!]
    \centering
    \includegraphics[width=\textwidth]{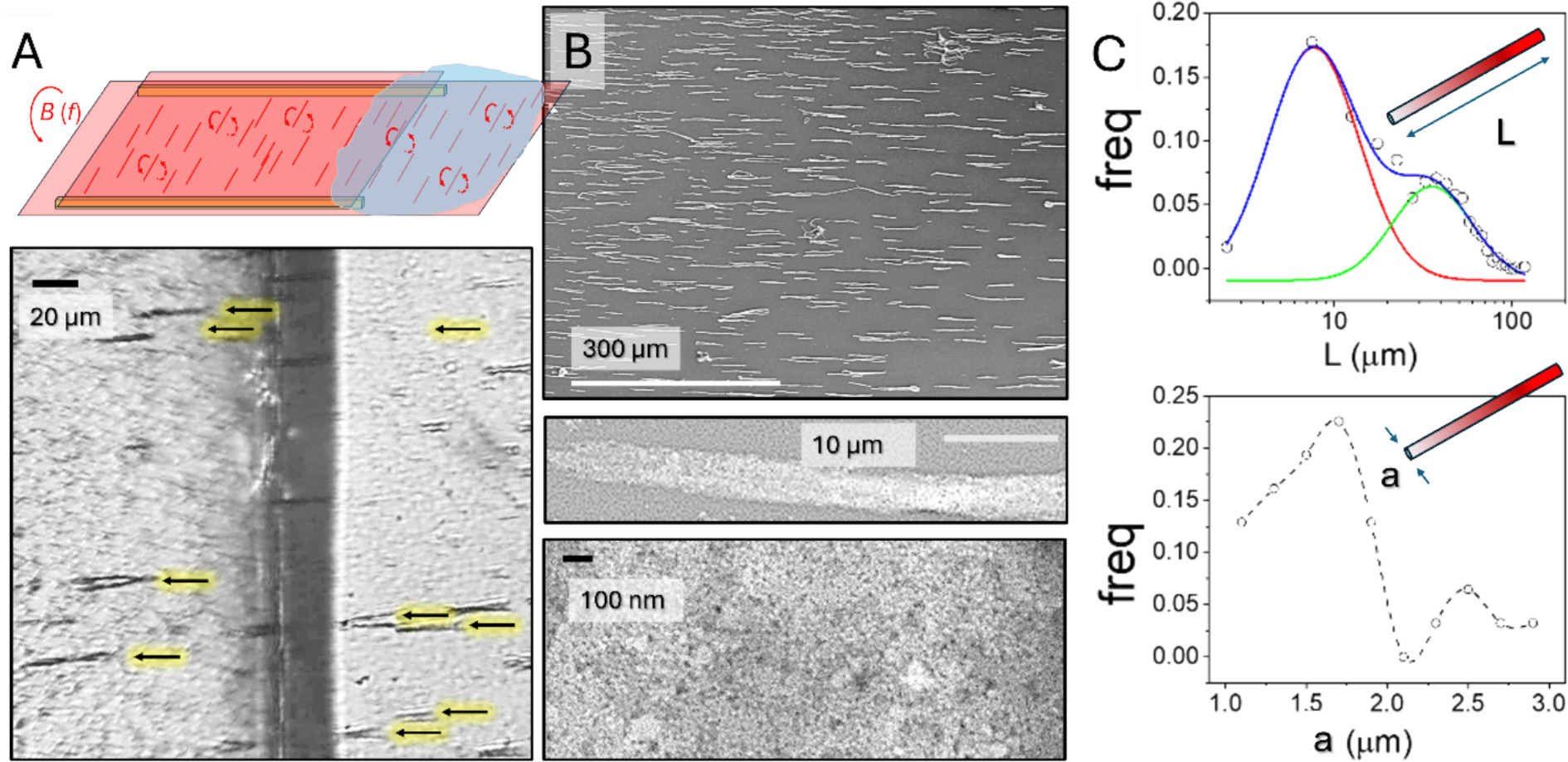}
    \caption{A) (top) The scheme illustrates the strategy for extracting filaments from the fabrication chamber
    by exploiting their coupled rotation and translation under a rotating magnetic field applied in the plane
    perpendicular to the substrate. (bottom) Images showing the driven transport of magnetic filaments under
    a rotating field ($B_z = 0.9$ mT, $B_x = 2.5$ mT, $f = 1.0$ Hz) and at a salt concentration of [NaCl] = 70.0 mM.
    Black arrows highlight the directed motion of the filaments as they move out through the open edge of the
    chamber, covering approximately 20 µm in 1.2 seconds. The thick lines running vertically through the
    center of the images mark the edge of the glass coverslip that bounds the chamber. B) SEM images of the
    magnetic filaments at different magnifications after extraction from the fabrication chamber and subsequent
    drying. C) Length, $L$, and diameter, $a$, distributions of the filaments collected using the procedure described.}
    \label{fig:Fig6}
\end{figure}

The images obtained by scanning electron microscopy (SEM), along with subsequent
analysis using the ImageJ software, allow for the determination of the size of the filaments
extracted from the chamber (Figure 6C). When the filaments are fabricated in the
chamber formed with the double-sided adhesive tape, the distribution of measured
filament lengths, $L$, can be deconvoluted into two clearly distinct size populations. The
first shows a peak around 7 $\mu$m and mainly consists of short filaments that did not reach
the full height of the chamber during their formation. The second displays a peak near 40
$\mu$m, which approximately corresponds to the height of the formation chamber. The
presence of filaments longer than the chamber height is attributed to possible aggregation
during the extraction process, likely driven by dipolar interactions induced by the applied
magnetic field. High-magnification images of the filaments (Figure 6B) reveal that there
is apparently no discernible structural order in the arrangement of the MNFs forming the
filaments.

After field exposure, the assembled filaments can be carefully extracted and re-dispersed
in a different solution to evaluate their post-field integrity and functional behavior. Once
collected using a micropipette and transferred into an SDS solution (5mM), to minimize
adsorption onto the substrate, the extracted filaments can be manipulated again through
the application of external fields. When the extracted filaments are subjected to a rotating
magnetic field, applied in the plane of the substrate, their motion is governed by both the
applied field and the viscoelastic properties of the surrounding medium \cite{Helgesen1990, Berret2016}. At low
Reynolds numbers, where inertial effects are negligible, the magnetic torque generated
by an externally rotating magnetic field is exactly balanced by the opposing viscous
torque. Under these conditions, at sufficiently low rotation frequencies of the magnetic
field, the filament rotates synchronously with the field. It maintains a constant angular
velocity equal to that of the field, with a steady phase lag between the direction of the
applied field and the orientation of the induced magnetic moment on the filament.
However, as the field frequency increases and surpasses a critical threshold,
corresponding to a phase lag of 90 degrees between the magnetic field and the magnetic
moment, the filament transitions into an asynchronous regime. In this regime, the
magnetic torque is no longer sufficient to overcome the increasing viscous resistance
required to maintain synchronous rotation. Consequently, the filament's angular velocity
falls behind that of the rotating field. The phase difference between the field and the
filament increases continuously over time, and the net magnetic torque periodically
reverses direction within each rotation cycle. This results in a back-and-forth, oscillatory
angular motion rather than continuous rotation. This dynamic is characteristic of the
asynchronous regime and is illustrated in Figure S8. The study of synchronous and
asynchronous dynamics in magnetic filaments is particularly relevant for characterizing
their magnetic susceptibility \cite{Helgesen1990, Berret2016}. It also suggests potential applications, such as
micro-scale mixing or transport devices.

\subsection{Size-controlled magnetic filaments made up of L-DOPA-coated negatively charged
MNFs}

The positive charge of the uncoated MNFs promotes their adsorption onto the glass substrate, which can hinder the manipulability and controllability of the resulting structures. To overcome this limitation, we followed the described protocol to coat the nanoparticles with L-DOPA, thereby rendering them negatively charged due to deprotonated carboxylate groups. In TEM images, the L-DOPA coating appears as a low-contrast organic layer surrounding the nanoflowers, while maintaining a defined interparticle spacing (see Figure 1S F). We then repeated the same experimental protocol using these L-DOPA-coated nanoparticles (MNFs@L-DOPA). Under a constant and homogeneous magnetic field oriented along the vertical direction, the field-induced assembly process is qualitatively similar to that observed for positively charged particles; however, the relevant salt concentration window is shifted to significantly lower values, spanning 0--20 mM. At very low electrolyte concentrations, below 5 mM, no microstructures capable of persisting in the absence of the magnetic field are formed. As in the positively charged system, the addition of salt promotes the formation of dense colloidal phases that act as nucleation seeds for field-induced assembly. Accordingly, salt concentrations of 5 mM or higher are required before filamentous structures retain their elongated morphology once the field is switched off. At salt concentrations above 20 mM, electrolyte screening promotes the formation of permanent aggregates prior to field application, as well as their adsorption onto the glass substrate, thereby preventing controlled field-induced assembly.

Unlike their positively charged counterparts, the resulting field-induced structures are not anchored to the substrate and, in the absence of the magnetic field, lose their collective orientation and positional order, becoming randomly distributed and oriented, while lying and floating on the confining plane (see Figure 7A). As a result, the formed filaments can be easily lifted or detached by the applied field, reflecting their low tendency to adsorb onto the glass. When the rotating field is applied in a plane perpendicular to the substrate, the reduced friction with the lower surface renders transport mechanisms based on rotating magnetic fields much less effective (compare Figures 5D and 7B). The absence of a stable anchoring point suppresses the rotation--translation coupling responsible for walking or crawling motion along the substrate, making directed transport inefficient or entirely absent. Consequently, the magnetic extraction strategy developed for anchored filaments is no longer applicable in this case. Instead, because the filaments are not attached to the substrate, they can be readily removed from the formation chamber by simply aspirating the suspension with a pipette. When subjected to a rotating magnetic field applied parallel to the substrate plane, these filaments rotate predominantly around their center of mass rather than around a fixed anchoring point (Figure 7C).

\begin{figure}[h!]
    \centering
    \includegraphics[width=\textwidth]{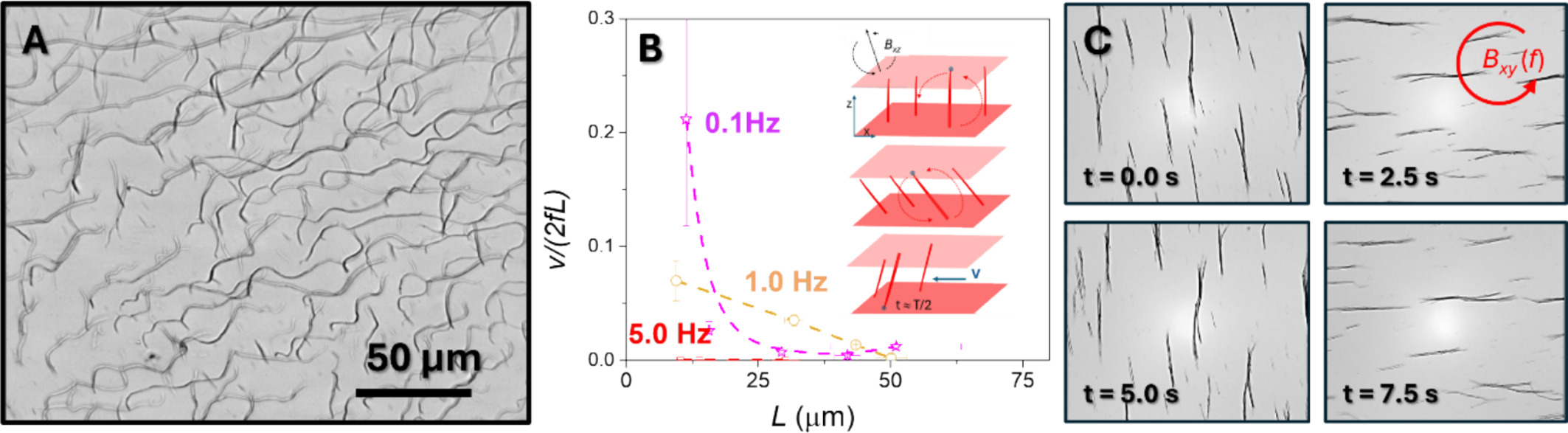}
    \caption{Suspension of L-DOPA coated MNFs in 10 mM NaCl. A) after removal of a magnetic field ($B_z
    = 0.9$ mT) applied for 300 s. B) Dimensionless velocity as a function of filament length for different field
    frequencies, under a circular rotating field applied perpendicular to the horizontal substrate ($B_z = B_x = 3.3$
    mT, $\beta = 0$). The averaged values and associated errors result from the analysis of at least three different
    filaments of similar length monitored under the same conditions. C) Time-lapse image sequence showing
    microfilaments rotating synchronously about their center of mass with the applied magnetic field ($B_x = B_y
    = 1.1$ mT, $f_x = f_y = 0.1$ Hz).}
    \label{fig:Fig7}
\end{figure}

The incorporation of L-DOPA coating provides a versatile strategy to introduce chemical
functionality while preserving the ability of the particles to self-assemble into higher-
order structures. Despite the additional surface layer, L-DOPA-functionalized
nanoflowers retain their capacity to undergo field-driven assembly into microfilaments,
indicating that the coating does not suppress the magnetic interactions required for
collective organization. In addition, L-DOPA provides a chemically active interface rich
in catechol and amine groups, which can promote secondary interactions and enable the
attachment or encapsulation of molecular species. This opens the possibility of combining
structural control with chemical functionality in a single system, without altering the
underlying assembly mechanism. Furthermore, the L-DOPA-functionalized assemblies
preserve their mechanical integrity and dynamic response under external magnetic fields.
The ability of these structures to undergo actuation after functionalization indicates that
L-DOPA does not hinder their responsiveness but rather offers an alternative route toward
stabilizing and functionalizing magnetically actuated architectures. Such robustness is
essential for extending these systems beyond proof-of-concept assemblies toward
functional materials.

\section{Conclusions}

In this work, we have established a versatile and fully template-free strategy for the controlled self-assembly of magnetic nanoflowers (MNFs) into reconfigurable
micropillars and microfilaments. By systematically tuning ionic strength and applying externally controlled magnetic fields, we achieved precise control over assembly pathways, enabling the formation of both reversible, liquid-like columnar structures and irreversible, solid-like micropillars. The interplay between electrostatic screening, magnetic dipolar interactions arising from permanent magnetic moments, and spatial confinement governs phase separation into liquid-like and gas-like domains at moderate salt concentrations, while higher ionic strengths promote stable aggregates. External magnetic fields accelerate and direct both processes, enabling dynamic control over structural integrity and actuation.

Our central hypothesis is that fundamental interparticle interactions---when properly balanced through ionic strength modulation and magnetic torque---are sufficient to drive programmable, scaffold-free assembly into functional micromagnetic architectures. Based on this concept, we demonstrate that stable, dynamic cilia-like actuators can be generated without predefined templates or patterned substrates. We further introduce torque-induced peeling of anchored micropillars to yield fairly monodisperse filaments, which can be extracted via coupled rotation--translation strategies. Additionally, we show that chemical functionalization through L-polydopamine coating preserves collective magnetic behavior, enabling multifunctional assemblies that integrate chemical reactivity with programmable magnetic motion.

In contrast to previous reports relying on patterned substrates, permanent matrices, or predominantly irreversible aggregation routes \cite{Ni2023, Kwon2024, Bharti2015, Belardi2011, Vilfan2010, Doyle2002, Nie2010, Fiser2015, Wang2019, Luo2019, Dreyfus2005, Kim2022, Tajuelo2016, Cebers2016, Mateos2019, Alphandery2012, Alexandridis2025, Zamani2022, Wang2003, Golovanov2013, Vereda2007, Kralj2015, FrkaPetesic2011, Fresnais2008, Biswal2003, Magdanz2020, Bereczk2017, Yusoff2021, Xu2019, Zhukov2004, Chiriac2011}, our approach provides reversible control over assembly pathways, tunable transitions between liquid-like and solid-like states, and integrated actuation and transport within a single adaptable platform. Moreover, while field-induced phase separation and electrolyte-driven dense-phase formation in magnetic colloids have been previously reported (e.g., references \cite{Bacri1994, Bacri1989, Dubois2000, Cousin2001, Raboisson2020, Erdmanis2017}), those studies primarily focused on phase behavior, droplet dynamics, or rotating-field-induced deformation. In contrast, we combine a systematic exploration of electrolyte concentration and magnetic field configuration with a detailed analysis of actuation mechanisms and reversibility in self-assembled architectures emerging from dense ferromagnetic nanoparticle phases. This integrated perspective enables programmable transitions between reversible and irreversible states, controlled anchoring, torque-induced peeling, and rotation--translation coupling, which extend beyond previously described phase-separation phenomena. Furthermore, the integration of efficient magnetic hyperthermia performance (large AC hysteresis loop areas, tunable coercivity, and high specific absorption rates) with field-driven structural organization establishes a multifunctional platform that couples biomedical heating capability with dynamic self- assembly. While magnetic hyperthermia and field-induced structuring have been extensively studied independently, their combination within a single, reconfigurable, template-free architecture remains comparatively unexplored.

Future efforts will focus on understanding the frequency-dependent dynamics of oscillatory filaments to optimize flow generation, mixing efficiency, and transport precision in confined environments. Extending magnetic field configurations to access alternative geometries---such as membranes, honeycomb lattices, or foams \cite{Muller2014, Osterman2009, Camacho2025, Martin2013}---could broaden structural diversity and functionality. Integration into closed microfluidic platforms and exploration under complex flow conditions will be critical for practical implementation. Moreover, given that phase segregation and complex assembly behaviors have been reported for other nanoparticle--salt systems \cite{Bacri1994, Dubois2000}, we anticipate that this template-free strategy can be generalized to a wide range of colloidal building blocks and ionic environments.

\balance

\bibliography{biblio}
\bibliographystyle{rsc}

\end{document}